\documentclass[a4paper,12pt]{article}

\usepackage[utf8]{inputenc}
\usepackage[T1]{fontenc}
\usepackage{amsmath}
\usepackage{placeins}
\usepackage{graphicx}
\usepackage[symbol]{footmisc}

\usepackage[a4paper,margin=2cm]{geometry}
\usepackage[backend=biber,sorting=none,style=numeric]{biblatex}
\title{Probing Geometry of Ion-Ion Collisions with Roman Pot Detectors}
\author{ 
Janusz J. Chwastowski, Krzysztof Cieśla, 
Rafa\l{} Staszewski\footnote{correspondig author: \texttt{rafal.staszewski@ifj.edu.pl}} \\[2ex]
Institute of Nuclear Physics 
Polish Academy of Sciences,\\
Radzikowskiego 152, 31--342 Kraków, Poland
\\[2ex]
Piotr Babiarz\footnote{graduated from AGH-UST}\\[2ex]
AGH University of Science and Technology,\\
Al. Mickiewicza 30, 30-059 Kraków, Poland
}

\date{\today}

\newcommand{\plotwidth}{0.49\columnwidth}

\usepackage{xspace}
\newcommand{\SZ}[1][]{\ensuremath{\sum Z_{#1}}\xspace}
\newcommand{\SA}[1][]{\ensuremath{\sum A_{#1}}\xspace}
\newcommand{\Npart}{\ensuremath{N_\text{part}\xspace}}

\begin{document}

\maketitle
\hrule
\begin{abstract}
A possible use of forward proton tagging detectors at the LHC in the context of heavy ion interactions is discussed. 
It is shown that signals registered in such detectors are sensitive to the geometry of the collision.
A method of the impact parameter and the collision asymmetry reconstruction on the event-by-event basis is proposed and its performance is studied with 
Monte Carlo simulations. 
\end{abstract}
\vspace{1ex}
\hrule
\vspace{2em}

\newcommand{\plotplaceholder}[1]{
\begin{figure}[htb]
    \centering
    \tikz{\draw[fill=black!5!white] (0,0) rectangle (5,3);}
    \caption{#1}
    \label{fig:#1}
\end{figure}
}

\section{Introduction}

The Large Hadron Collider \cite{lhc} is able to accelerate and collide various beams.
The machine has successfully run in the proton--proton, proton--lead, lead--lead modes.
In addition, several test runs involving xenon ions have been performed.

In an interaction of two ions at the LHC energies, the size and the evolution of the created medium depend on the collision geometry.
In the present paper, the impact parameter, $b$, will be used for describing this geometry.
Events characterised by $b \approx 0$ are called central collisions while those with $b$ close to $2R$ are called the peripheral ones.

\begin{figure}[htb]
    \centering
    \includegraphics[width=\plotwidth]{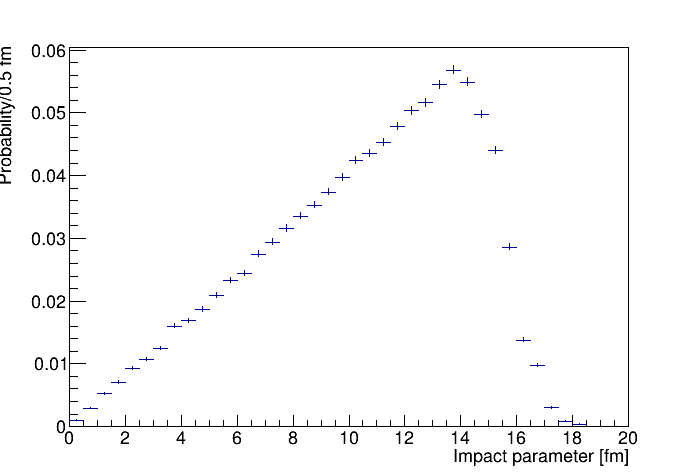}
    \caption{Impact parameter probability distribution.}
    \label{fig:b-dpmjet}
\end{figure}

From simple geometric considerations, one expects that the probability density for having a collision with a certain $b$ value grows linearly with $b$ between $b=0$ and $b=2R$ and for $b>2R$ immediately drops to zero (for rigid spheres).
The distribution of the $b$ value calculated from 100000 lead--lead collisions at the nucleon--nucleon centre-of-mass energy $\sqrt{s}_{NN} = 5.04$~TeV produced using DPMJET-III Monte Carlo generator~\cite{dpmjet} is shown in Figure \ref{fig:b-dpmjet}. 
For $b<2R$\footnote{$R \approx 7$~fm for lead ions.}, the DPMJET distribution follows simple expectations.
For $b>2R$, one observes a steep but smooth drop due to a more realistic treatment of the effects related to the nucleus edge.
This is an example of a more general phenomenon originating from a complex structure of the colliding particles -- the impact parameter is not sufficient to describe the full geometry of the interaction.

An alternative description of the geometry can be obtained considering the structure of a nucleus and  describing the heavy ion collision in terms of the number of nucleons taking part in the interaction, \Npart, or the number of binary collisions, $N_{coll}$.
Then, one expects that peripheral processes lead on average to smaller values of \Npart or $N_{coll}$ than those observed for the central ones.

\begin{figure}[htb]
    \centering
    \includegraphics[width=\plotwidth]{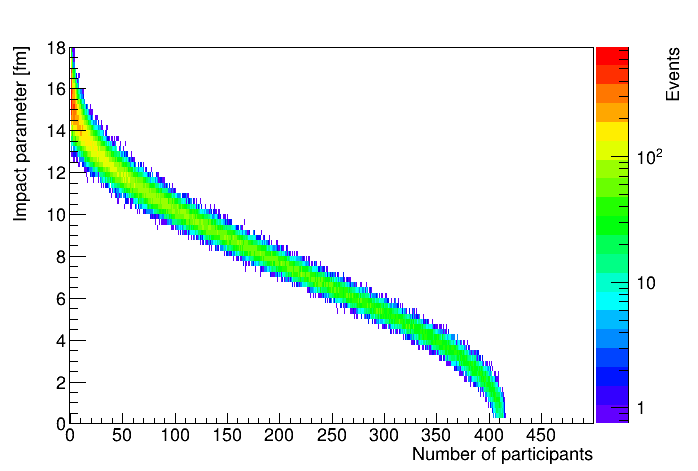}
    \caption{Correlation of the impact parameter and \Npart.}
    \label{fig:cbNpart}
\end{figure}

Figure \ref{fig:cbNpart} presents the relation between \Npart and $b$ as predicted by DPMJET-III Monte Carlo.
It shows a very strong and anticipated correlation.
The non-negligible width of the correlation indicates the importance of fluctuations of the nucleus shape in the initial state of the collision.
It also confirms that the majority of collisions are of peripheral nature.
A qualitatively similar picture can be seen (not shown here) in the case of the $N_{coll}$ dependence on $b$. 

Since neither $b$ nor \Npart nor $N_{coll}$ are directly measurable, the centrality of an event is usually experimentally defined on the basis of some easily measured observable sensitive to the geometry of the collision. 
For example, one often uses the multiplicity of centrally produced hadrons or the energy measured in forward calorimeters -- see \cite{alice-b} and \cite{atlas-b} for description of the methods. 
These methods rely on the Glauber model and are sensitive to the details of the experimental apparatus. 

In \cite{tarafdar}, the authors considered an alternative method of estimating \Npart in gold--gold collisions at RHIC.
They proposed a new system of Cherenkov detectors capable of measuring the majority of spectator fragments scattered at small angles.
The proposed method is based on an observation that for a full acceptance detector system, the measurement relies only on the energy conservation and therefore is model independent.
The measurement of the debris was also discussed in \cite{hera} in the context of the potential $eA$ interactions at HERA.

In the present paper, a different approach is considered. 
It is investigated  whether detectors that do not provide a full coverage, and hence can register the spectator fragments only partially, can deliver any valuable information about the collision geometry. 
The problem is studied for the forward proton detectors already operating at the LHC.

The paper is organised as follows. Section \ref{sec:fpdet} introduces the considered experimental apparatus.
This is followed by the discussion of the apparatus acceptance in Sec. \ref{sec:accept}. 
The impact parameter dependence on the registered debris mass and atomic numbers is described in Sec.~\ref{sec:depen}.
A study of the asymmetry of the geometry is discussed in Sec. \ref{sec:asymmetry} which is followed by a Summary.

\FloatBarrier

\section{Forward proton detectors}
\label{sec:fpdet}

In the following, the ATLAS Forward Proton (AFP) detectors \cite{afptdr, Aad:2020glb} are considered as the registering devices.
These detectors measure protons emitted or scattered at very small angles and thus escaping registration in the ATLAS main detector.
Such protons traverse the magnetic lattice of the accelerator, which serves as a magnetic spectrometer. 
The detector uses the Roman pots technique, which allows for precise positioning of the active parts in the immediate vicinity of the beam.

It is quite obvious that the detector acceptance depends on the properties of the LHC machine.
In addition, the position of the detector, quantified by the distance between the detector active part and the beam, plays a crucial role.
A typical distance is about 2 -- 3~mm which covers around 15 widths of the beam at the detector position and about 0.5~mm of the dead space due to the experimental infrastructure.
The AFP detectors take data during standard operation of the LHC.

The four Roman pot stations are positioned symmetrically with respect to the ATLAS interaction point at the distances of about $|z| = 205$~m and $|z|=217$~m.
The stations allow horizontal, i.e. in the accelerator plane, motion of the Roman pots.  
Each station contains a silicon tracker made of four precise 3D pixel planes.
The planes are tilted w.r.t. the $x$-axis (horizontal direction) and staggered in the $y$-axis (vertical direction).
The resulting spatial resolution of the scattered proton track measurement is about 10~$\mu$m and 30~$\mu$m in the horizontal and vertical direction, respectively.
The detector area as seen by the scattered protons is about 16~mm by 20~mm.
The scattered proton energy is obtained indirectly using unfolding of the trajectory measurement leading to the reconstruction resolution better than 10~GeV \cite{afprecresol}.
The stations at $|z=217$~m contain also time-of-flight counters providing the resolution of the order of 20 -- 30~ps.
These counters can be used for rejecting combinatorial background rejection originating from large pile-up present in LHC proton-proton runs, see \cite{Staszewski:2019yek} for more details.
Since such backgrounds are not relevant to the present study, the use of time-of-flight detectors is not considered here.

The important variables describing a scattered proton are its transverse momentum, $p_T$, relative energy loss, $\xi = (E_{beam}-E')/E_{beam}$ where $E_{beam}$ is the beam energy and $E'$ denotes the scattered proton energy and the azimuthal angle $\varphi$.
The scattered proton is registered by the AFP detectors with high acceptance if its relative energy loss is within the interval of $(0.02, 0.12)$ and the transverse momentum $p_T < 3$~GeV \cite{afpacceptance}.

The considerations presented below required simulation of the particle/debris transport through the magnetic lattice of the LHC.
These calculations were performed using the MaD-X  \cite{madx} code and the optics files describing the accelerator settings used in lead--lead runs.
For simplicity, the particles were transported to the middle point between the AFP stations i.e. up to 211~m from the ATLAS interaction point.
More details about the simulation can be found in \cite{acta}.

\FloatBarrier

\section{Acceptance for nuclear fragments}
\label{sec:accept}

The discussed method of centrality estimation is based on a simple fact that the collision geometry affects both the number of participants and the number of spectators.
Therefore, the measurement of the spectators should provide information about this geometry.
In the case of the AFP detectors, a direct measurement of neither the total multiplicity of nucleons nor the total energy of the produced fragments is possible%
\footnote{%
The measurement of the total energy would be of interest since all spectator nucleons have, to some approximation, energy equal to their energy before the interaction.
The approach used for energy reconstruction used in the proton--proton case cannot be used for nuclear debris because of an additional degree of freedom -- the unknown charge of the given ion.
}.
However, since the silicon detectors are sensitive to the amount of ionisation caused by passing particles, the measurement of the spectator electric charge could be possible with appropriately tuned sensors.

\begin{figure}[htb]
    \centering
    \includegraphics[width=0.49\textwidth]{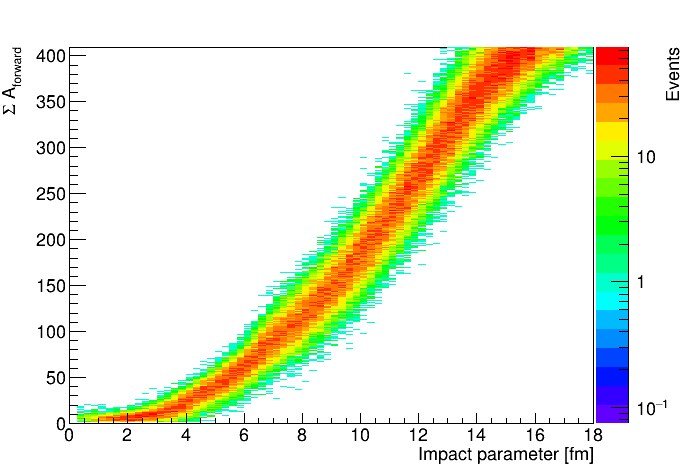}
    \includegraphics[width=0.49\textwidth]{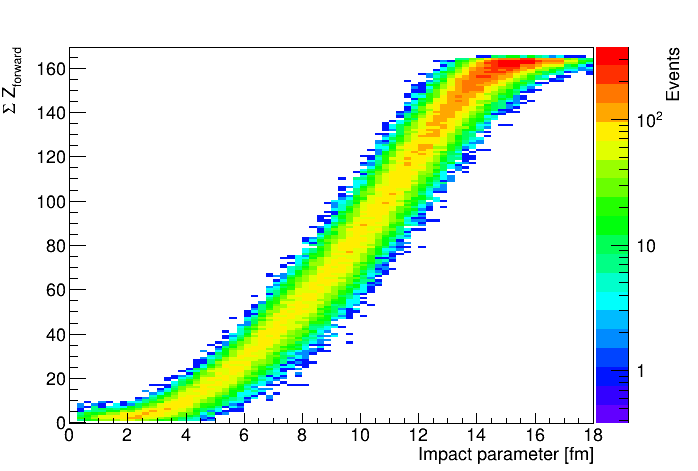}

    \caption{Left: Correlation of the sum of the mass numbers of 
    the nuclear debris emitted in the forward direction, 
    \SA,  and the impact parameter.
     Right: Correlation of the sum of the atomic numbers of 
    the nuclear debris emitted in the forward direction, 
    \SZ, and the impact parameter.}
    \label{fig:SumAtotvsb}
    \label{fig:SumZtotvsb}
\end{figure}

In the present paper, the following scenarios will be considered:
\begin{itemize}
\item an ideal case -- the measurement of nucleon multiplicity, \SA,
\item a more realistic case -- the measurement of the charge multiplicity, 
\SZ, of the forward moving spectators.
\end{itemize}
It is educative to compare correlations between the above observables and the impact parameter, see Figure \ref{fig:SumAtotvsb}.
While there are some differences between these two cases, the correlation patterns are actually quite similar.
The one involving \SZ is a little wider but this effect is not large.
Therefore, one may expect that the measurement of \SZ should provide information about the impact parameter comparable to that delivered by the measurement of \SA.

The next step of the present analysis was devoted to the determination of the AFP response to the nuclear debris originating from the non-interacting, forward moving system of spectators.
The calculations, performed in a model independent way, followed the lines of  an earlier study \cite{acta} and considered all known nuclei. 
At first it was checked that the life times of produced fragments allow for their potential registration at the AFP positions before decaying.
Later, using the Mad-X~\cite{madx} description of the LHC, the transport of these nuclei was simulated. 
It was assumed that the detectors operate at a distance of 3~mm from the beam.

\begin{figure}[htb]
    \centering
    \includegraphics[width=\textwidth]{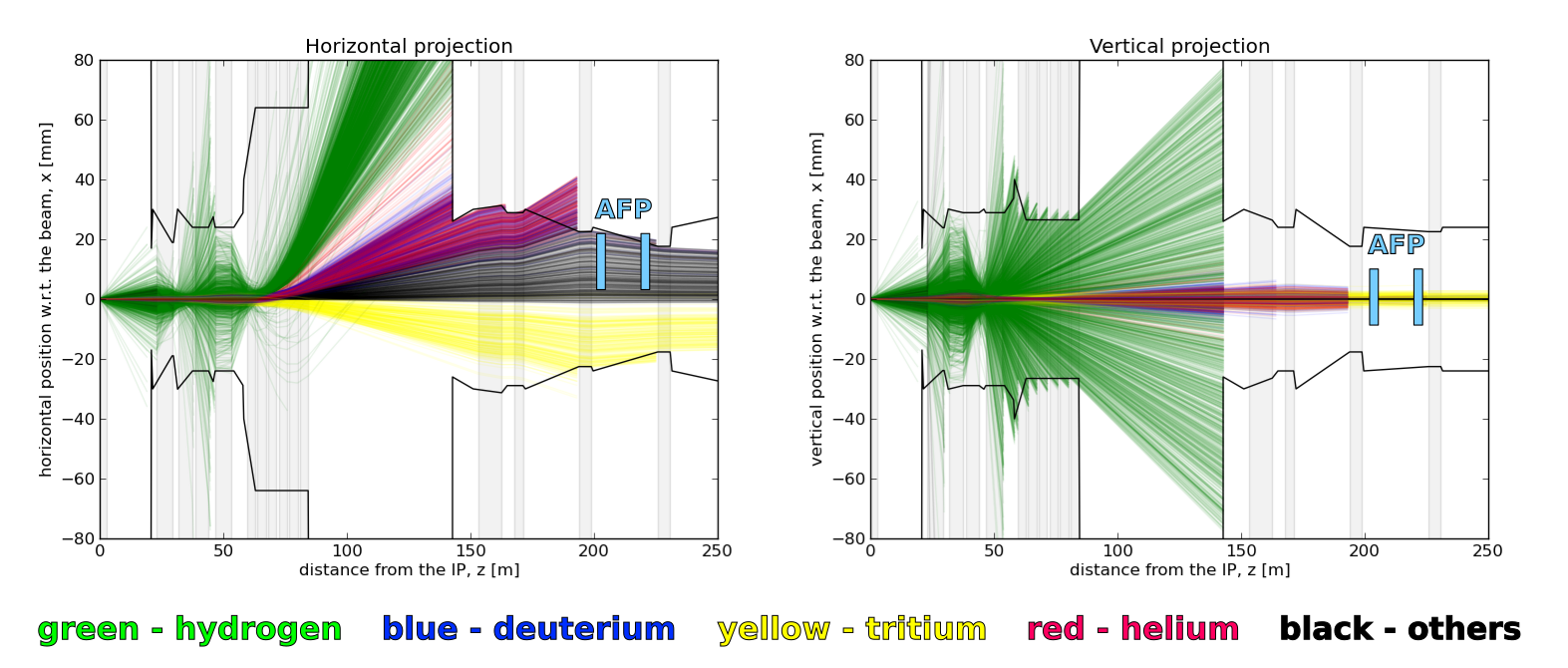}
    \caption{Projections of the trajectories of nuclear 
    fragments in $(x, z)$-plane (left panel) and 
    $(y, z)$-plane (right panel). 
    }
    \label{fig:HI_trajectories}
\end{figure}

The projections of the trajectories of the ions in $(x, z)$ and $(y, z)$ planes are presented in Fig.~\ref{fig:HI_trajectories}.
The accelerator magnetic lattice filters out spectator protons (marked with the green lines), deuterium (blue), tritium (yellow) and helium (red).
The spectator neutrons, not shown in the plot, are neglected in the present analysis.

\begin{figure}[htb]
    \centering
    \includegraphics[width=0.49\columnwidth]{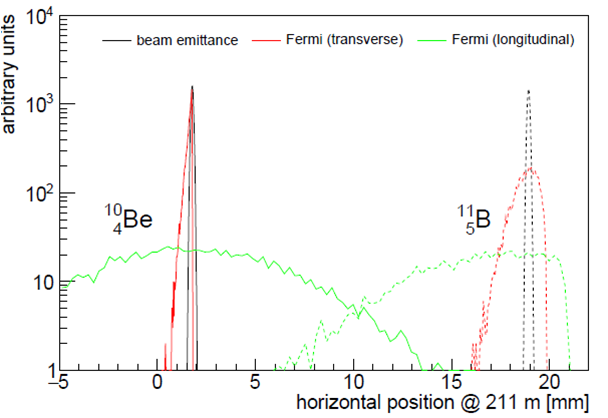}
    \includegraphics[width=0.49\columnwidth]{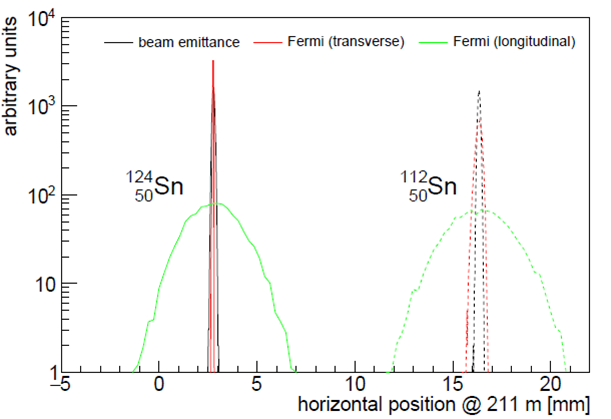}
    \caption{Effects of the beam emittance and Fermi motion on
    the ion position at the AFP detector. Upper panel -- beryllium and boron, lower panel -- tin. From
    \cite{acta}.}
    \label{fig:smearingSn}
\end{figure}

The momentum of a nuclear fragment emerging from an ion--ion collision is predominantly driven by its mass number.
However, it is influenced by the Fermi motion of the nucleons belonging to the considered fragment and effects related to the finite emittance%
\footnote{The beam emittance is a measure of the spread of the beam particles in the position--momentum phase space: $(x, p_x)$ or $(y, p_y)$.}
of the colliding beams.
Without these effects, nuclear fragments of a given type would always be observed in the detector at the same position.
The Fermi motion and the finite emittance introduce a spread of this position, which is illustrated in Figure \ref{fig:smearingSn} for beryllium, boron and tin ions. 

The beam emittance plays a very small role leading to a minuscule broadening of the position distribution and was therefore neglected in the following analysis.
On the contrary, the Fermi motion strongly affects the ion horizontal position. 
Its longitudinal component has a much stronger impact because its influence is magnified by a large value of the Lorentz factor of the colliding beams.
For collisions at $\sqrt{s}_{NN} = 5.02$~TeV, $\gamma \approx 2700$ leading to the potential smearing of the nucleon momentum up to nearly 1400~GeV and hence to the enhanced smearing of the horizontal position of the fragment.
In the present study, this influence was found much weaker for heavier ions as an effect of a non-correlated Fermi motion of different nucleons assumed.

\begin{figure}[htb]
    \centering
    \includegraphics[width=0.49\textwidth,page=2]{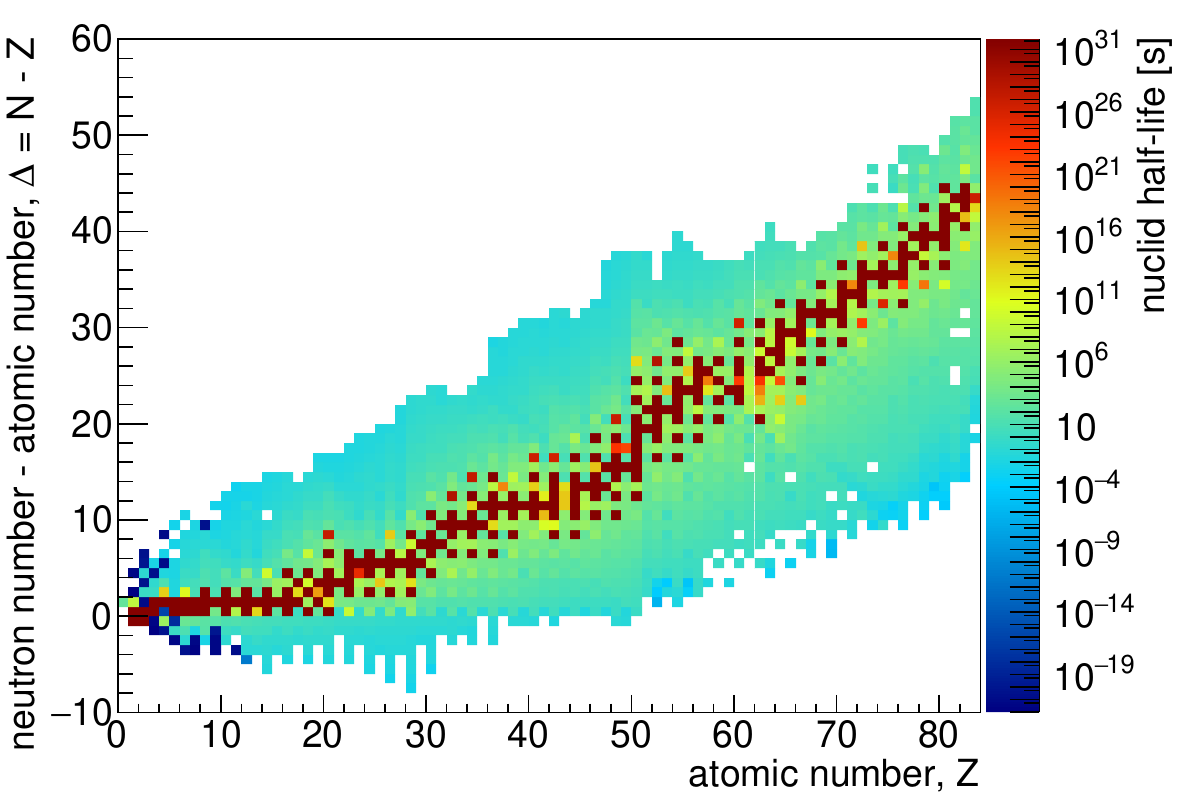}
    \includegraphics[width=0.49\textwidth,page=5]{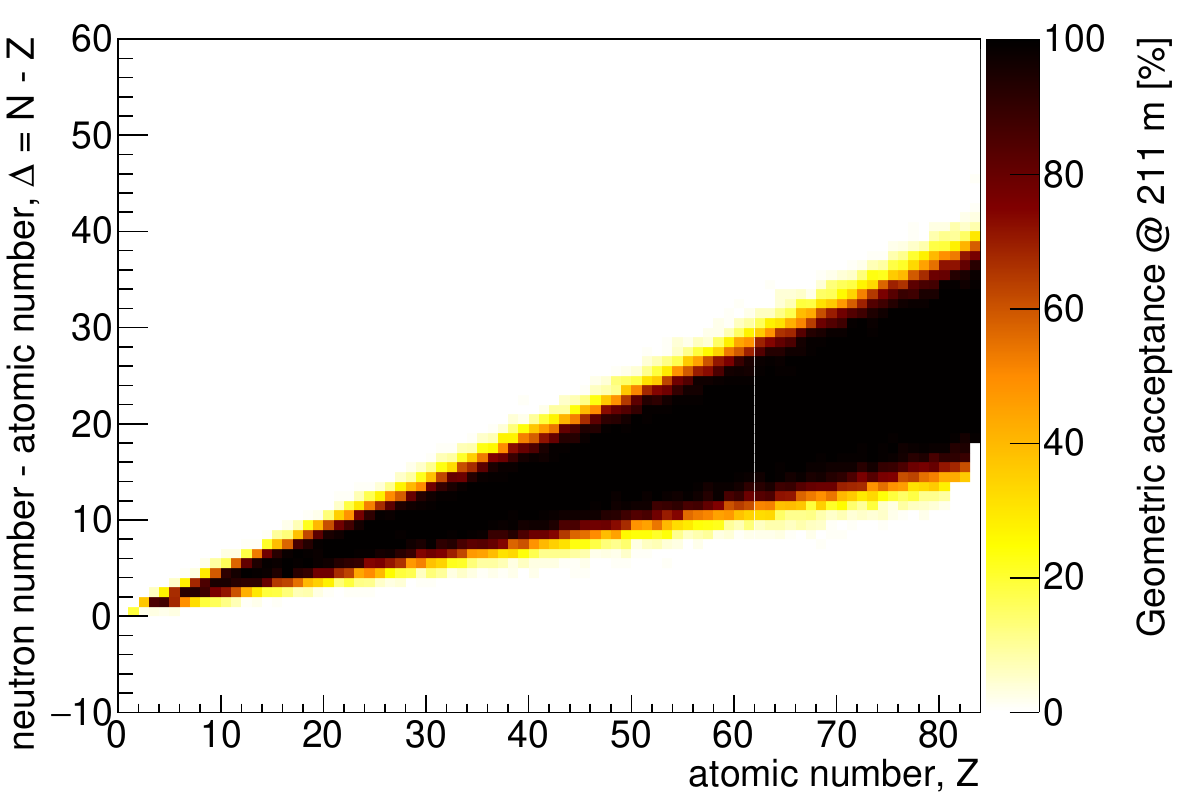}
    \caption{
    Left: half-life times of known nuclids.
    Right: the AFP detector acceptance calculated including the beam emittance and Fermi motion effects.
    From \cite{acta}.}
    \label{fig:smearedAcceptance}
\end{figure}

The impact of the above discussed effects on the AFP capability to register various nuclei is summarised in Figure~\ref{fig:smearedAcceptance} showing the detector acceptance as a function of $(Z, \Delta)$, where $\Delta$, calculated as $\Delta = A-2\cdot Z$, is the net number of neutrons (the surplus/deficit of neutrons with respect to the protons) in a nucleus. 
The AFP acceptance is smeared but the region of high  acceptance value is clearly visible for a broad range of nuclei.

\FloatBarrier

\section{Dependence of centrality on the registered fragments}
\label{sec:depen}

The considered forward proton detectors cannot observe all possibly created nuclear fragments.
This raises a question of how much the limited acceptance affects the possibility to estimate the collision geometry.
This type of analysis cannot be carried out in a model-independent way and the following results are again based on the DPMJET-III simulation with the transport of the fragments calculated using the Mad-X.

\begin{figure}[h!]
    \centering
    \includegraphics[width=0.49\columnwidth]{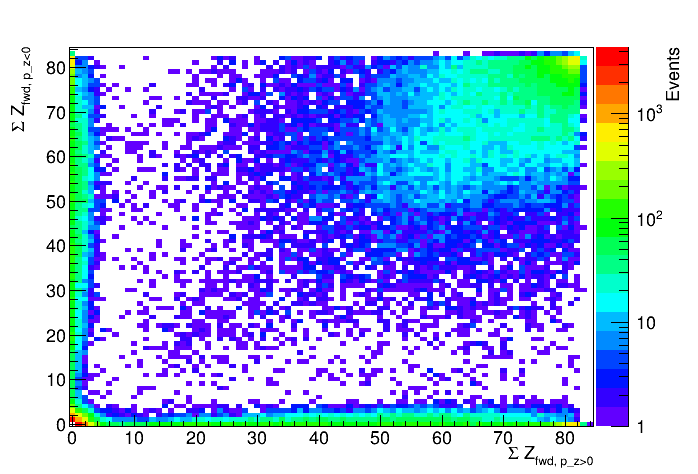}
    \caption{Correlation of the total charge the nuclear debris recorded by the forward proton detector on the two sides.}
    \label{fig:MeasSumZ_fvsSumZ_b}
\end{figure}

Figure \ref{fig:MeasSumZ_fvsSumZ_b} shows the correlation between the total charge of debris registered by the detectors located at both beams: \SZ[p_z>0] vs. \SZ[p_z < 0].
One can distinguish three classes of events with different signatures.
The first class contains events for which large values of \SZ are observed on both sides.
The second one contains events with an asymmetric configuration:
a large value of \SZ on one side and a very small \SZ on the opposite side.
The third class includes events in which both detectors register small total charge.
A correlation between the signals on both sides is observed for the first class of events.
The \SZ values separating the classes can be estimated looking at the projection of Fig. \ref{fig:MeasSumZ_fvsSumZ_b} on one of the axes.
A minimum around $\SZ = 10$ is observed and this value was selected and is used in the following for defining the no-tag, single-tag and double-tag events.

\begin{figure}[htb]
    \centering
    \includegraphics[width=.49\textwidth]{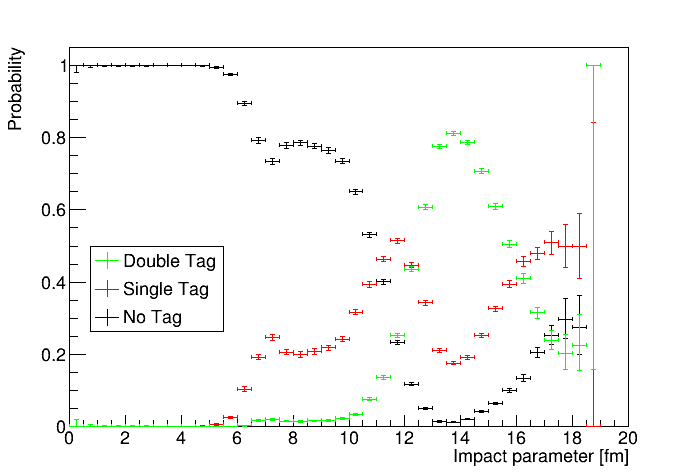}
    \includegraphics[width=.49\textwidth]{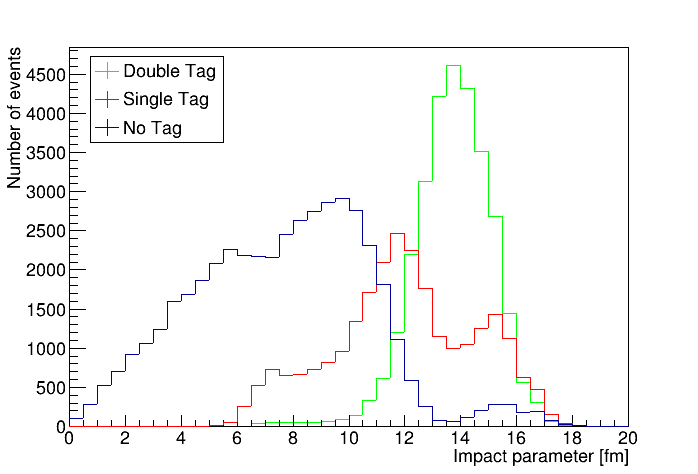}
    \caption{
    Left: Probability of observing different classes as a function of impact parameter.
    Right: distribution of impact parameter for events with different classes.
    }
    \label{fig:effi}
\end{figure}

Naturally, the signature of an event depends on the collision geometry, see Figure \ref{fig:effi} (left).
Most central events have no tag in the forward detectors. Single-tag events are most likely for $b\approx 12$~fm and $b>16$~fm.
For $b$ around 14 fm, the double-tag signature becomes the most likely class.
One may also ask a question if the signature alone provides information about the event centrality.
This can be deduced from Figure \ref{fig:effi} (right), showing the distribution of $b$ for the three classes.
One can observe that no-tag events are predominantly central ones, double-tag events are rather peripheral.
Single-tag events are on average in between no- and double-tag ones but their distribution is quite wide and have a considerable coverage with the other classes.

\begin{figure}[htb]
    \centering
    \includegraphics[width=0.49\textwidth]{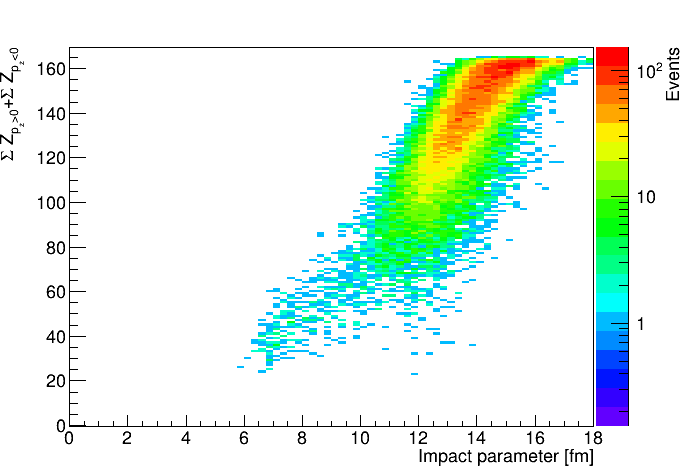}
    \includegraphics[width=0.49\textwidth]{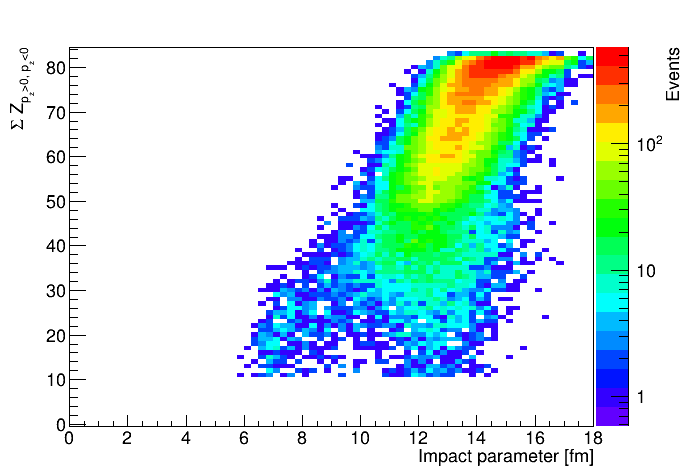}
    \caption{Correlation of the sum of charges of the nuclear debris recorded by  the forward detectors, \SZ, and the impact parameter.}
    \label{fig:MeasSumZtotvsb}
\end{figure}

In order to check the possible sensitivity of the registered signal to the collision geometry, a correlation between the total charge and the actual impact parameter of the collision was investigated. 
Figure~\ref{fig:MeasSumZtotvsb} shows this correlation for the single- and double-tag classes (note different vertical ranges in both plots).
Comparing this result to Figure \ref{fig:SumZtotvsb} one can immediately observe that limiting the acceptance results in an increased width of the correlation.

Summarising, it is clear that the considered detectors, while able to register only a part of the produced spectator fragments and measuring only their charge, can provide useful information about the geometry of the collision.
The next step is to understand the possible performance of the method, namely the resolution with which the impact parameter can be reconstructed.

The resolution of the impact parameter reconstruction is driven mainly by the randomness in the formation of the nuclear fragments that eventually reach the detectors.
This effect is responsible for the non-zero width of the correlation presented in Figure \ref{fig:SumAtotvsb}.
It is further enhanced by the limited acceptance of the detectors, which can be seen in Figure \ref{fig:MeasSumZtotvsb}.
Extracting the width along the $b$ direction and interpreting it as a possible measurement resolution at a given $b$ allows a comparison of different methods and different assumptions.

\begin{figure}[htb]
    \centering
    \includegraphics[width=\plotwidth]{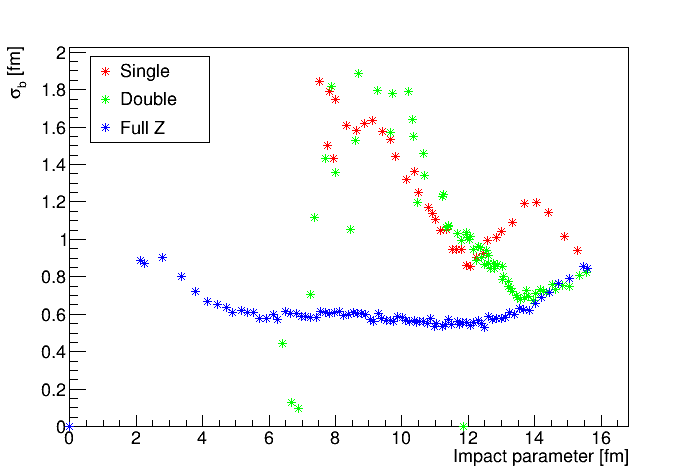}
    \caption{Resolution of impact parameter reconstruction based on spectator fragments for single- and double-tag events as well for as an ideal situation assuming full acceptance forward detectors.}
    \label{fig:sigma}
\end{figure}

Figure \ref{fig:sigma} presents the resolution of the impact parameter reconstruction obtained for single- and double-tag events.
For central events with $b < 8$~fm, the results for double-tag class are heavily influenced by the limited statistics of the generated events as well as for the ideal case of full acceptance detectors.
In all other cases, the resolution between 1 and 2 fm is observed.
These values are of the order of 10\% of the range in which $b$ varies in lead--lead collisions.
For events with $b<12$~fm, single-tag events offer a better resolution, while for $b>12$~fm double-tag events lead to a more precise estimation.
For $b>14$~fm, the resolution for double-tag events is close to the ideal case.
It is worth pointing out that the widths of the correlations in Figure \ref{fig:MeasSumZtotvsb} along the \SZ axes are of the order of $\sigma(\SZ) \approx 10$.
This value can be interpreted as the maximum magnitude of resolution in the \SZ that would not dominate the $b$ resolution.

\FloatBarrier
\section{Collision asymmetry}
\label{sec:asymmetry}

\begin{figure}[htb]
    \centering
    \includegraphics[width=0.49\textwidth,page=2]{figures/asym.pdf}
    \includegraphics[width=0.49\textwidth,page=1]{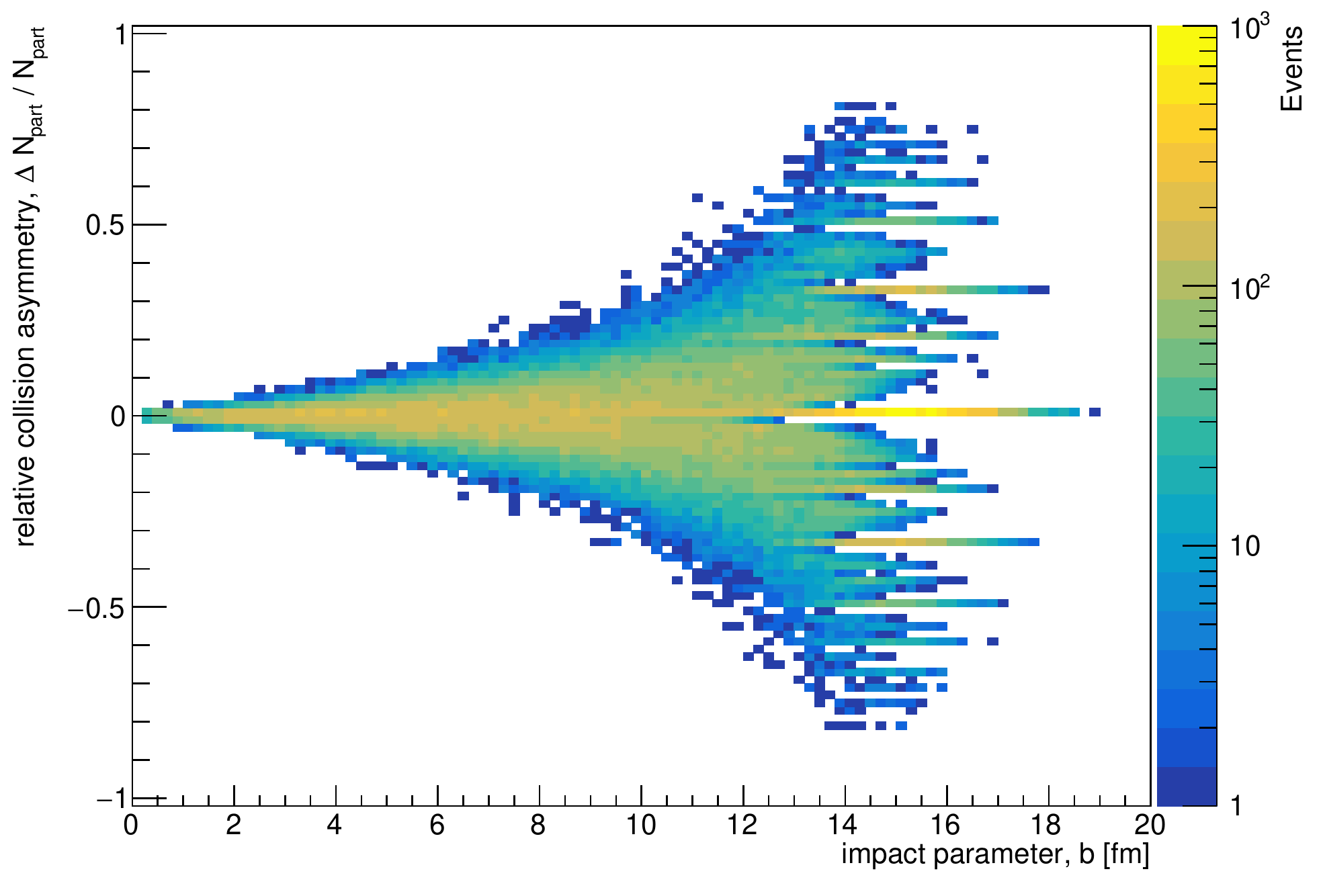}
    \caption{Distributions of absolute (left panel) and relative (right panel) collision asymmetry as a function of the impact parameter.}
    \label{fig:asym_all}
\end{figure}

All previous considerations assumed that there is one parameter describing the geometry of the heavy-ion collision.
This assumption is true for colliding symmetric objects with smooth internal structure.
In the case of realistic interactions, important asymmetries can be present already in the initial state.
They can be quantified, for example, by the difference in the number of participating nucleons from each of the ions.
Figure \ref{fig:asym_all} shows how the distribution of the asymmetry depends on the impact parameter.
The biggest absolute differences, $\Delta \Npart$, can reach even a few tens of nucleons and occur for medium centralities with $b$ around 8~fm. 
However, when one considers relative asymmetries, i.e. $\Delta \Npart/\Npart$, the highest values are observed for the most peripheral collisions, where the total number of participants is the smallest.

\begin{figure}[htb]
    \centering
    \includegraphics[width=0.49\textwidth,page=6]{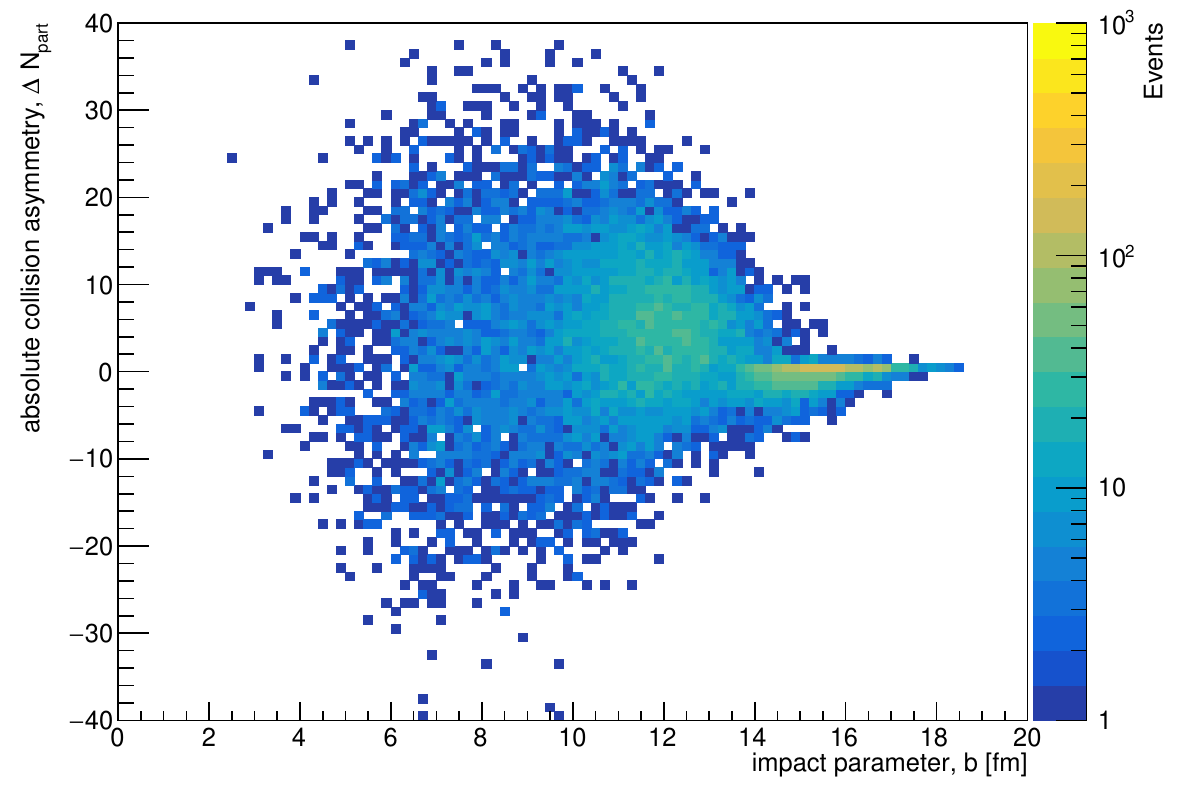}
    \includegraphics[width=0.49\textwidth,page=8]{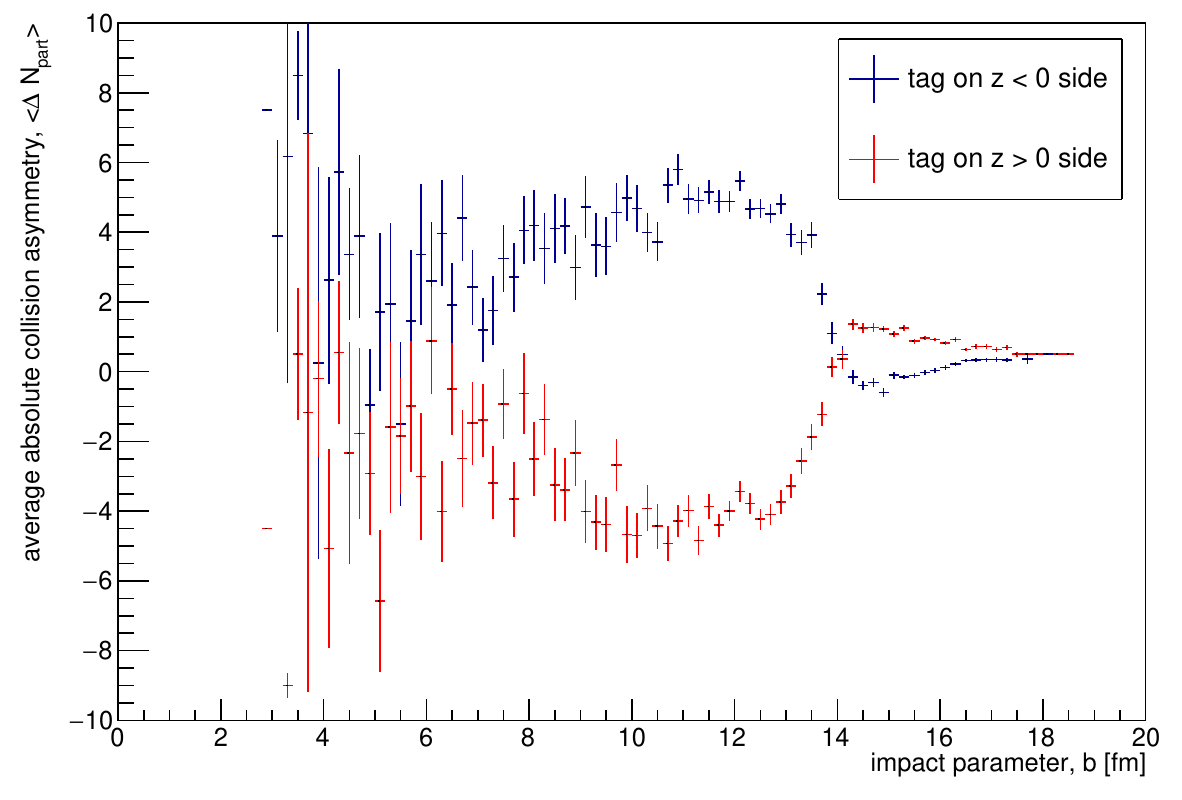}
    \caption{Left: distribution of collision asymmetry as a function of the impact parameter for single-tag events with the tag on the $z<0$ side.
    Right: average collision asymmetry as a function of the impact parameter for single-tag events with the tag on each side.}
    \label{fig:asym_single}
\end{figure}

Since forward proton detectors can measure fragments originating from both ions, they could provide information not only about the centrality but also about the asymmetry of the collision.
The way this information can be extracted will depend on the event signature.
For single-tag events, the information about the asymmetry can be extracted from the information on which side the event was tagged.
This can be observed in Figure \ref{fig:asym_single}, where the asymmetry distribution as a function of $b$ is shown for events tagged on one side.
It is interesting to observe that the shape of the asymmetry distribution changes with the impact parameter.
In fact, even a change of the sign of the mean value of the asymmetry is observed.
This is a non-trivial consequence of the way the spectator fragments are formed and of the limited acceptance of the detectors.

\begin{figure}[htb]
    \centering
    \includegraphics[width=0.49\textwidth,page=9]{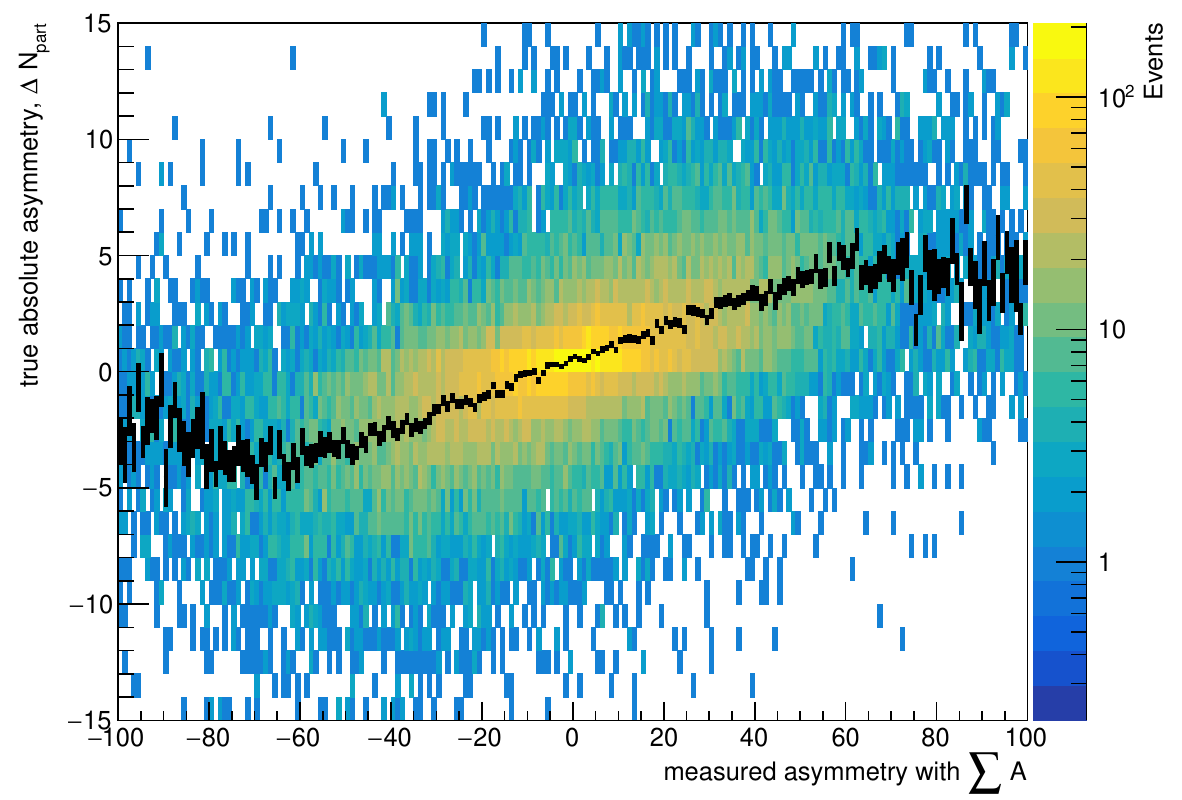}
    \includegraphics[width=0.49\textwidth,page=10]{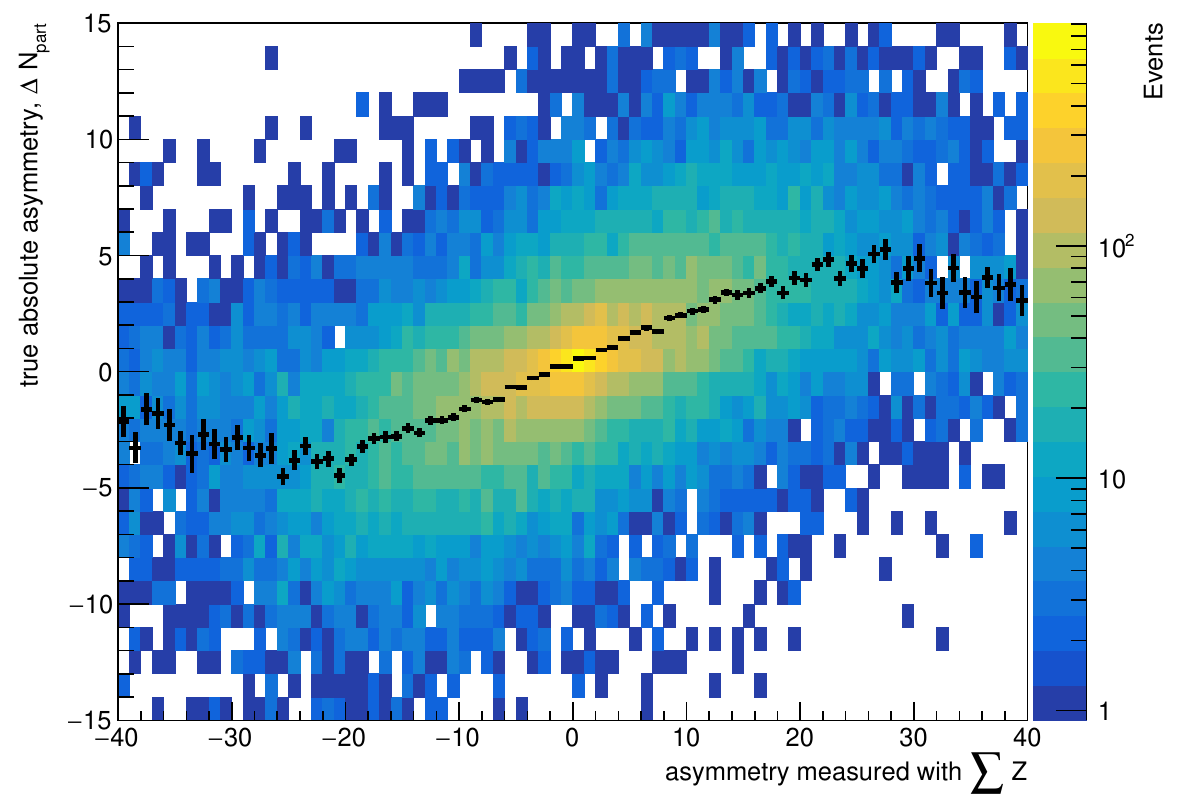}
    \caption{Correlation between the true asymmetry and the asymmetry measured with \SA (left) or \SZ (right).}
    \label{fig:asym_double}
\end{figure}

For double-tag events, one may apply a more direct approach and calculate the experimental asymmetry using the measured \SA values on both sides.
Figure \ref{fig:asym_double} presents the correlation between the true and the measured asymmetry of the collision.
While the distribution is rather wide, this method does provide some sensitivity%
\footnote{The correlation coefficient between the true and the measured asymmetry was found to be close to $0.5$.}
to the true asymmetry.

\FloatBarrier
\section{Summary and Conclusions}

A possibility of the application of the forward proton detectors in heavy ion collision at the LHC was investigated. 
The impact parameter determination on the event-by-event basis was studied using DPMJET-III generated Monte Carlo events and Mad-X calculations of particle trajectories in the accelerator magnetic fields.
The calculations demonstrated that the existing detectors (ATLAS Forward Proton detectors were used as an example) have a significant acceptance to a wide range of known nuclei.
It was found that the Fermi motion of the nucleons belonging to a fragment very strongly impacts the position measured in the forward detectors while the beam emittance plays a negligible role.
In simulations, the detectors were used to tag forward emitted debris on one or both sides of the collision.  
The performed analysis suggests that the charge measurement of the debris  delivers 1 -- 2~fm precision of the impact parameter reconstruction.
It was also shown that the method can be used for a rough determination of the collision asymmetry.

Two facts have to be stressed. 
First, the above discussed results depend on the properties of the accelerator magnetic lattice, and hence the calculations have to be repeated for each case separately.
Second, the present results rely on the physics model of the spectator system fragmentation used.
Therefore, the present work can be extended towards the understanding of the uncertainty of this modelling and how this translates into a possible uncertainty on the collision geometry reconstruction.

\section*{Acknowledgements}
This work was supported in part by Polish Ministry of Science and Higher Education grant no. DIR/WK/2016/13 and Polish National Science Centre grant no. 2015/19/B/ST2/00989.

\printbibliography

\end{document}